\documentstyle[bo99,epsfig]{article}

\title{Low-luminosity AGN and Normal Galaxies}
\author{A.~Ptak}
\affil{Carnegie Mellon University, Dept. of Physics, 5000 Forbes Ave.,
Pittsburgh, PA 15213}

\begin{document}

\maketitle

\begin{abstract}
Low-luminosity AGN (with $L_X < 1 \times 10^{42} \rm \ ergs \ s^{-1}$) 
far outnumber ordinary AGN, and
are therefore perhaps more relevant to our understanding of AGN phenomena
and the relationship between AGN and host galaxies.  Many normal galaxies
harbor LINER and starburst nuclei, which, together with LLAGN, are a class
of ``low-activity'' galaxies that have a number of surprisingly similar X-ray
characteristics, despite their heterogenous optical classification. 
This strongly supports the hypothesis of an AGN-starburst connection.
Further, X-ray observations of normal galaxies without starburst or AGN-like
activity in their nuclei offer opportunities to study populations of X-ray
binaries, HII regions, and warm or hot ISM under different conditions than
is often the case in the Milky Way.  The results of recent
X-ray observations of these types of galaxies are reviewed, and what we hope to
learn about both nearby and high redshift galaxies of each type from
observations with forthcoming and planned satellites is discussed.

\end{abstract}

\section{Introduction}
Since active galactic nuclei (AGN) comprise only a small fraction of 
all galaxies, it is more
relevant for our general understanding of galactic processes to observe normal 
galaxies.  With the availability of {\it imaged} X-ray observations, galaxies
not necessarily dominated by a point source in their nucleus have now been
studied in detail.  This review will give an overview of these results, with
an emphasis on galaxies exhibiting moderate or low amounts of activity in
their nuclei.

In this paper, a low-luminosity AGN (LLAGN) is considered to be a galaxy with
Seyfert-like optical spectra and an X-ray luminosity less than $10^{42} \rm \
ergs \ s^{-1}$.
A starburst galaxy is a galaxy in which evidence of enhanced
star-formation rates are observed, particularly in kpc-sized nuclear regions.
Most often this is observed by the presence of HII-region like optical
emission lines, however other signs include high IR luminosities (i.e.,
$L_{IR} > 10^{10} L_\odot$; c.f., Telesco 1988).  In LINER galaxies the
optical emission lines are observed with line diagnostic ratios that
differ from both starburst and LLAGN ratios.
  The most likely scenarios for the
ionization flux in LINERs are a LLAGN (albeit with different physical
properties than LLAGN with Seyfert-like spectra), shocks (Dopita et
al. 1996) and hot stars (c.f., Shields 1992).  Interestingly,
15\% of LINERs exhibit {\it broad} $H\alpha$ emission (Ho, Filippenko,
Sargent, \& Chen 1997). These ``LINER 1'' galaxies are almost certainly LLAGN,
and
the fact that the ratio of LINER 1 to LINER 2 galaxies is similar
to the fraction of Seyfert 1/Seyfert 2 galaxies is suggestive that most LINERs
are indeed LLAGN.

\subsection{Demographics}
In order to properly access the prevalence of LLAGN, LINER and starburst
activity, it is necessary to perform a careful survey of nearby galaxies in
which the galactic light is subtracted from the nuclear spectrum in order for
the often-weak optical emissions line to be detected.  Such as survey was
recently completed by Ho, Filippenko \& Sargent (1997).  The demographics of
these types of activity is shown in Figure 1.  Clearly, ``low-activity''
galaxies (starburst, LINER, and LLAGN, and transition nuclei containing both
LLAGN and starburst emission), 86\% of the total, dominate over ``normal'' 
galaxies.
Furthermore, ``normal'' AGN only comprise $\sim 10\%$ of all galaxies (Ho,
Filippenko, \& Sargent 1997).  This emphasizes the importance of studying these
types of galaxies: the usual state of affairs is for a galaxy to exhibit
starburst or LLAGN activity (or both) in its nucleus, and most AGN in the
local universe are in a low-luminosity state.
\begin{figure}
\centerline{\psfig{file=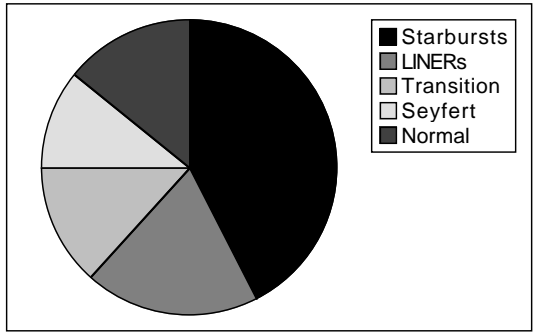, width=7cm}}
\caption[]{Demographics of activity in nearby galaxies.}
\end{figure}
 
\section{Normal Galaxies}
In elliptical/early-type
galaxies the X-ray emission is dominated by hot gas (in some cases similar to
group or cluster of galaxies X-ray emission, with a King-like radial surface
brightness profile, occasionally including a cooling flow), with a temperature
of $0.8-1.0$ keV (Trinchieri, Fabbiano, \& Kim 1998).  Some early-type
galaxies also contain a contribution from a LLAGN (see Di Matteo, et al.;
these proceedings) and 
low-mass X-ray binaries (LMXB, Irwin \& Bregman, 1998).
\begin{figure}
\centerline{\epsfysize=5cm \epsfbox[300 450 375 650]
{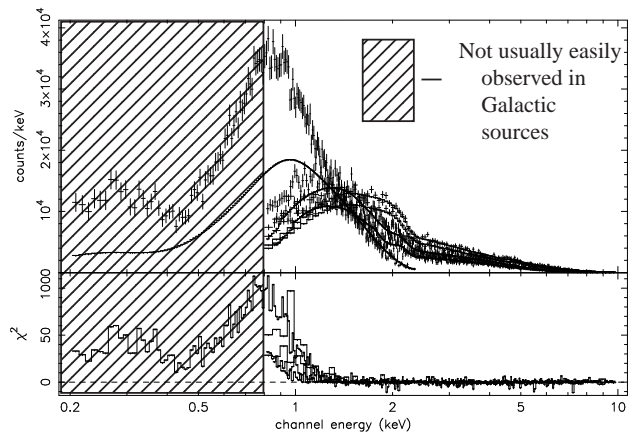}}
\caption[]{ASCA + PSPC spectra from the bulge of M31. Figure taken from Irwin
\& Bregman (1999), with a shaded area added to delineate the region typically
absorbed in Galactic X-ray sources.}
\end{figure}

The X-ray emission of spiral galaxies is generally dominated by X-ray
binaries (Fabbiano 1989).  Spiral arms often contain HII regions, particularly
regions of local density enhancements (i.e., knots).  These HII regions are
often X-ray bright as a result of the products of enhanced star formation:
hot stars, supernovae (SN) and supernova remnants (SNR), high-mass X-ray
binaries and black hole candidates (BHC) (c.f. NGC 1313 in Colbert et
al. 1997).  Not surprisingly, ``bluer'' galaxies, which tend to be later-type
galaxies with higher star-formation rate (with the blue colors being a result
of a higher proportion of massive stars than is observed in galaxies dominated
by older stellar populations) are X-ray bright (Fabbiano, Feigelson, \&
Zamorani 1982).  There is also a strong correlation between X-ray luminosity
and 
IR luminosity, with the IR emission being produced by dust that has been 
heated by massive stars (David, Jones \& Foreman 1992; Green, Anderson \& Ward
1992).  In cases of very high star formation rates occurring near the edge of
galaxies, blow-outs can occur when the local pressure resulting from outflows
from hot stars and SN exceeds the ambient interstellar medium (ISM) pressure, 
as observed by ROSAT in galaxies such as NGC 55 (Schlegel et
al. 1997).  There is also X-ray evidence that normal galaxies (Cui et
al. 1996), including the
Milky Way, possess hot ($T \sim 10^6$K) ``halos'' or ``coronae'' (Spitzer
1956).  
%

Spiral galaxies also typically contain a bulge component, which may be
counterparts to elliptical galaxies.  A substantial fraction of the X-ray
sources found in the nearest spiral galaxy, M31, are concentrated in the
bulge.  These sources are presumably mostly LMXB associated with the older
stellar populations in globular clusters.  The X-ray spectrum of the
bulge region of M31 as observed by the ROSAT PSPC and ASCA (Irwin \& Bregman
1999) and BeppoSAX (Trinchieri et al. 1999).  This spectrum has a power-law
component consistent with X-ray binaries in our galaxies, plus a soft
component (see Figure 2) that is similar to that considered to be due to X-ray
binaries in early-type galaxies.  Note that it would be difficult to observe
this soft component in Galactic binaries since most lie in the disk of the
Milky Way and hence are highly absorbed.

Colbert and Mushotzky (1999) discuss a survey of X-ray sources in normal
galaxies with luminosities in excess of $1.3 \times 10^{38} \ \rm ergs \
s^{-1}$, the Eddington luminosity for accretion onto a solar-mass object.  
These intermediate-luminosity X-ray objects (IXO) have X-ray spectra consistent
with accretion onto black holes with masses of $\sim 10^{1-4} M_{\odot}$.
IXOs are typically significantly displaced
from the galactic centers, and hence they are not AGN.  This implies that
IXOs may be cases of ``intermediate'' mass black holes, and may be precursors
to some modern-day AGN (see also M82 below and in Ptak \& Griffiths 1999).
 
\section{Low-Activity Galaxies}
\subsection{Spectral Properties}
Despite their heterogeneous
optical classification, low-activity galaxies usually have a similar spectral
shape (Serlemitsos, Ptak \& Yaqoob 1996).
Specifically, in general these galaxies exhibit at least two spectral
components: a soft, thermal component with kT $\sim 0.7$ keV and a hard
component well-modeled by either a thermal bremsstrahlung with kT $\sim 5-10$
keV or a power-law with $\Gamma \sim 1.8$ (Ptak et al. 1999).  
The hard (soft) component is typically absorbed by a 
column density of $\sim 10^{22} \
\rm cm^{-2}$ (
$\sim 10^{20-21} \ \rm cm^{-2}$).  
The fact that the hard
component tends to be more spatially compact (see below) and absorbed than the
soft  
component implies that the hard component is emanating from further within
the galaxies (i.e., the nuclei).  The fact that starburst galaxies exhibit a
hard component, likely due accreting sources, and LINER and LLAGN galaxies
exhibit soft emission, likely due to starburst activity, strongly supports the
idea of a starburst/AGN connection. 
\begin{figure}
\centerline{\psfig{file=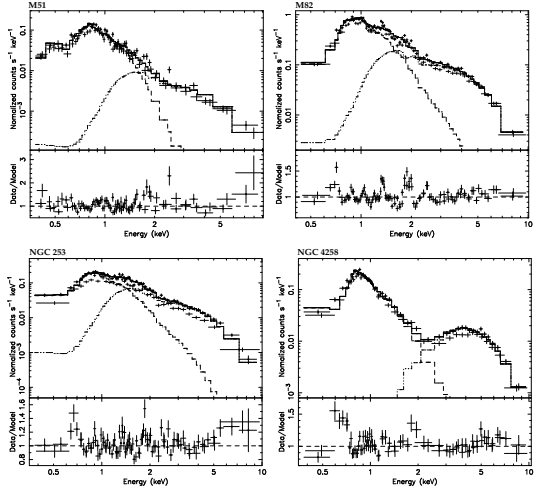, width=7cm}}
\caption[]{ASCA SIS spectra of M51 (starburst + LLAGN), M82 (starburst), NGC
253 (starburst) and NGC 4258 (LLAGN).}
\end{figure}

The luminosity of the hard component tends to be on the order of $10^{40-41} \
\rm ergs \ s^{-1}$ and the luminosity of the soft component $10^{39-40} \ \rm
ergs \ s^{-1}$, with the relative intensity varying from galaxy to galaxy (see
Figure 3).  
It is evidently rare for starburst activity, the
origin of the soft component, to achieve luminosities in excess of $10^{40-41}
\ \rm ergs \ s^{-1}$ (Halpern, Helfand, \& Moran 1995; however see
Moran, Lehnert \& Helfand 1999 for a counter-example in NGC 3256).
Accordingly, sources with X-ray luminosities $> 10^{41} \ \rm ergs \ s^{-1}$
(e.g., NGC 3998 and NGC 3147) only require a power-law 
component since the starburst component is overwhelmed.  Conversely, in Seyfert
2s where the AGN is highly absorbed, starburst emission is often observed
below 2 keV (see Turner et al. 1998).

The abundances inferred from the soft component tend to be 
sub-solar (on the order of 10$^{-1}$ solar).  This is surprising since
starburst emission is presumably the result of massive star evolution and
accordingly should be highly enriched, however many effects might be
contributing to this.  For example, it is probably too simple to be fitting
the 
multi-temperature starburst emission with a single component
(c.f., Breitschwerdt \& Komossa 1999; Dahlem, Weaver \& Heckman 1998), or
other sources of continuum may be present such as soft emission from X-ray
binaries (c.f., Figure 2). 
In brighter sources (see the residuals in Figure 3 and
Ptak et al. 1997), it
appears that there is a deficiency of Fe relative to $\alpha$-process elements
(e.g., Ne, Mg, Si, S, etc. produced in
massive stars), although the effect is diminished somewhat when more complex
models are invoked (see Dahlem, Weaver, \& Heckman 1999).
\subsection{Fe-K Emission}
Fe-K emission is an important diagnostic in AGN studies since its energy,
physical width, and equivalent width (EW) are functions of the physical
conditions 
in the accretion region.   
In the case of an obscured
nuclear region, the EW of an Fe line is expected to be greatly enhanced since
the direct continuum is diminished.  Although most low-activity galaxies are
too faint for Fe-K to detected, in several cases 
Fe-K {\it is} detected and is often
complex.  For example, the Fe-K EW in NGC 3147 (Ptak et al. 1996), NGC 1365
(Iyomoto et al. 1997), M51 (Terashima et al. 1998a), NGC 4736 (Roberts,
Warwick, \& Ohashi 1999) and NGC 1052 (Weaver et al. 1999) are high ($>$ 
100 eV)
relative to Seyfert 1 EW values ($\sim 100-200$ eV; Nandra et al. 1997),
consistent with an obscured nucleus.  In M81, the Fe-K line appears to contain
several components or may be broad, possibly due to an accretion disk
(Ishisaki et al 1996; Serlemitsos, Ptak \& Yaqoob, 1996). 
NGC 4579 (Terashima
et al. 1998b; Terashima et al. 2000ab) is a particularly interesting case where
the line was observed to be due to  {\it ionized} material in a 1995 ASCA
observation, but was observed to be due to neutral material in 1998.  
%

(Ionized) Fe-K emission was detected marginally by ASCA in M82 (Ptak et
al. 1997) and in M82 and NGC 253 significantly by 
BeppoSAX (Cappi et al. 1999).  This
emission stongly suggests that very hot gas (T $\sim 10^8$ K) is present in
these starburst galaxies.  However, the EW of the Fe-K lines is only on the
order of $100-200$ eV, considerably less than that expected from
solar-abundance hot gas (EW $\sim 600$ eV).  Again, the Fe abundance may be
depressed in these galaxies, however it is also likely that other sources of
continuum are present, diluting the thermal Fe-K emission.  

\subsection{Temporal Properties}
Low-activity galaxies tend to vary on long (months to year) time scales (c.f.,
M81 in Ishisaki et al. 1996; NGC 4579 in Serlemitsos, Ptak \& Yaqoob 1996), 
but not on short time scales as observed in Seyferts (Ptak et al. 1998; 
however note that M81 has been observed to vary at the 30\% level on day time
scales; Pellegrini et al. 2000).  
Suprisingly, some of the
most variable low-activity galaxies have been starbursts.  For example, the
nuclear source in the starburst NGC 3628 ``shut off'', varying by a factor of
$\sim 40$ (Dahlem, Heckman, \& Fabbiano 1995), and M82 has varied considerably
in the 2-10 keV bandpass (Ptak \& Griffiths 1999; Matsumoto \& Tsusru 1999;
Gruber et al., these
proceedings).  This implies that at least some of the contribution to the hard
component in starburst galaxies is due to accreting sources.   The lack of
variability on short times scales is demonstrated in Figure 4.  As argued in
Ptak et al. (1998), this marked break from the temporal behavior of Seyfert 1s
implies a large source extent for the X-ray producing regions in low-activity
galaxies.  While in some cases this might be due to a multiple sources of
X-ray emission, it may also be due to the prevelance of advection-dominated
accretion disks, in which the entire disk contributes to the X-ray emission,
as opposed to the ``$\alpha$''-disks in Seyferts in which the X-rays are most
likely produced by flares.  
On the other hand, the short-term variability
observed in M82 may be our first look at the hard X-ray light curve of an IXO,
assuming that the source of the 2-10 keV variability is the off-nuclear point
source observed by the ROSAT HRI (Collura et al. 1994).
\begin{figure}
\centerline{\psfig{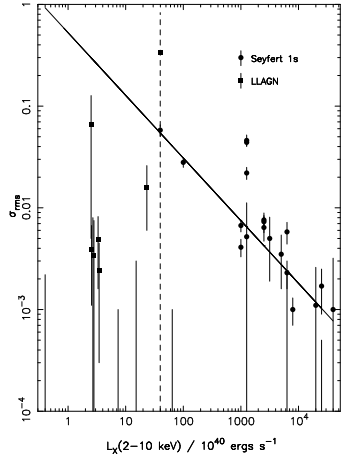}}
\caption[]{The trend of ``excess variance'', a measure of short-term
variability, with X-ray luminosity in Seyfert 1s and low-activity galaxies,
from Ptak et al. (1998).}
\end{figure}

\subsection{Spatial Properties}
Spatially most low-activity galaxies have
2-10 keV emission which is concentrated in the nucleus (see Ptak 1997, 
although the
statistics are limited in 
many cases) and are extended over kpc scales in addition to being concentrated
in multiple point sources.  This tends to be particularly true of starburst
galaxies (c.f.,
Read, Ponman, \& Strickland 1997; Dahlem, Weaver, \& Heckman 1998)
but these phenomena are also seen in LINERs and LLAGN (c.f., 
the LINER NGC 4594 in Fabbiano \& Juda 1997;
the LINER NGC 3079 in Pietsch, Trinchieri, \& Vogler 1998; 
the LLAGN M51 in Marston et al. 1995, 
and the LLAGN NGC 4258 in Cecil, Wilson, \& De Pree 
1998).  In one of the nearest starburst galaxies NGC 253 (at $\sim 2.5$ Mpc),
$\sim 73$ point sources (some of
which are background QSOs) have been detectedby ROSAT (Vogler \& Pietsch 1999;
see Figure 5), in 
addition to the extended emission associated with the nuclear region, disk,
and halo of NGC 253.  Interestingly, the point source distribution in NGC 253
is consistent with that of M31 (with the older-population buldge sources
removed) and M33 (Figure 6).  A similar result is observed more generally by
Roberts et al. (these proceedings), which provides motivation that a {\it
universal} 
luminosity distribution of X-ray binaries can be found, and the high-luminosity
end of which would be the IXOs.
\begin{figure}
\centerline{\psfig{file=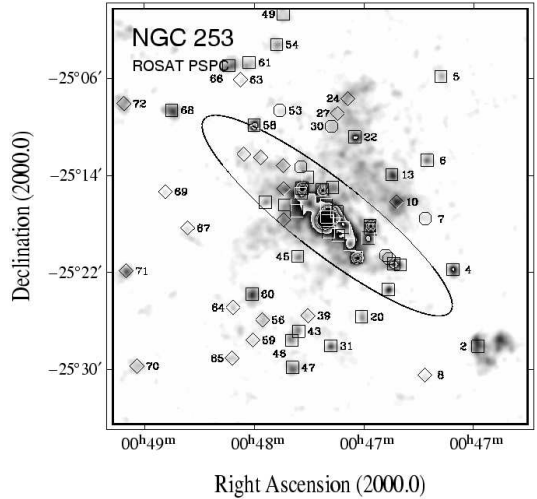, width=7cm}}
\caption[]{ROSAT PSPC image of NGC 253, from Vogler \& Pietsch (1999).}
\end{figure}
\begin{figure}
\centerline{\psfig{file=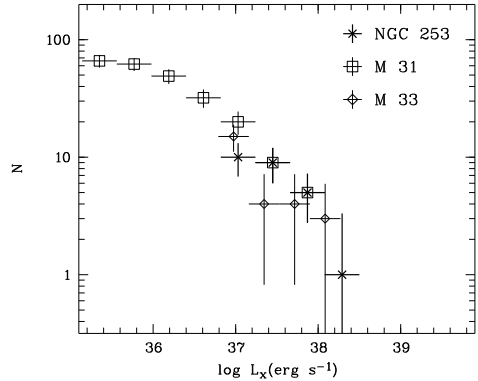, height=5cm}}
\caption[]{The luminosity distribution of point sources in NGC 253,
M33 and M31 (excluding buldge sources), from Vogler \& Pietsch (1999).}
\end{figure}

\section{Conclusions and Future Prospects}
Both ``normal'' galaxies and low-activities provide very rich data
sets for studying phenomena that is difficult to study in the Milky Way (i.e.,
due to extinction) and/or is not present in the Milky Way (i.e., starburst
regions with very high star formation rates resulting in superwinds).  The
X-ray emission occurs in both point sources, which tend to be the most
luminous known X-ray binaries and supernovae, and in complex diffuse emission
that is the result of heating of the ISM by hot star, supernovae and, in some
cases, superwind outflows.  The brightest X-ray binaries (IXOs) are black hole
candidates that potentially have masses on the order of $10^{1-4} \ 
M_{\odot}$ , intermediate to that of Galactic BHC and AGN.  Both starburst and
accretion emission tend to be observed whenever either type of activity is
present, strongly supporting the notion of a starburst/AGN connection.  There
appears to be a natural upper-limit to the luminosity of starburst processes
on the order of $10^{41} \rm \ ergs \ s^{-1}$, and so starburst emission is
usually overwhelmed by AGN emission that is not absorbed and exceeds $L_X \sim
10^{42} \ \rm ergs \ s^{-1}$.  Observed abundances tend to be absurdly low,
but that is almost certainly due to ``contamination'' of the continuum from
multiple temperature gas emission and unresolved point sources.  Chandra and
XMM will be able to resolve the starburst emission spatial and extract CCD
resolution spectra,
which will resolve this issue and allow abundance and temperature
enhancements to be observed in individual regions within the galaxies.

The X-ray properties of nearby normal and low-activity galaxies suggest that
they will contribute less $\sim 10\%$ of the X-ray background (c.f.,
Griffiths \& Padovani 1991; Yaqoob et al. 1995).  This fraction can
increase if these galaxies become harder and more luminous at earlier 
epochs, i.e., due to the enhanced starburst activity associated with the peak
in the star formation history of the universe at redshifts of $\sim 1-2$
(Hughes et al. 1998).  It may be possible to directly observe evolution in
low-activity galaxies with ultra-deep XMM surveys or with telescopes such as
XEUS that promise effective areas on the order of m$^2$ (see Figure 7).
\begin{figure}
\centerline{\psfig{file=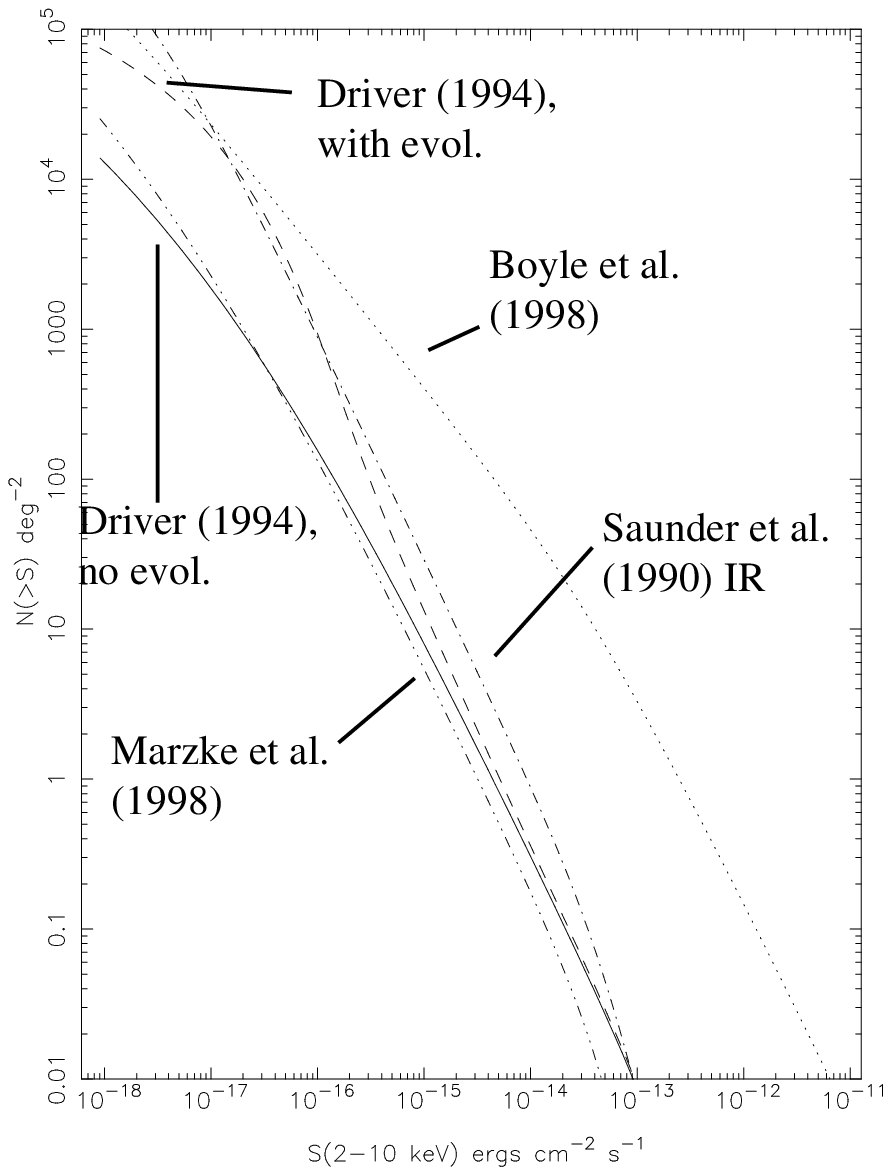, height=5.8cm}}
\caption[]{The expected logN-logS distributions in the 2-10 keV bandpass
obtained by converting the optical luminosity functions of Driver (1994) and
the IR luminosity function of Saunders et al. (1990) to the X-ray bandpass
using the optical and IR to X-ray correlations of David, Jones \& Forman
(1992) and Green, Anderson, \& Ward (1992), respectively.  Note that for the
mean power-law spectral slope of low-activity galaxies described in Ptak et
al. (1999) of $\Gamma \sim 1.7-1.8$, $F_{2-10 \ \rm keV} = F_{0.5-4.0 \ \rm
keV}$.}
\end{figure}

\end{document}